\documentclass[twocolumn]{aastex701}

\usepackage{amsmath,amssymb}
\usepackage{amsthm}
\usepackage{gensymb}

\begin{document}

\title{What are Switchbacks?}

\author[0000-0001-9570-5975]{Zesen Huang}
\affiliation{Department of Earth, Planetary, and Space Sciences, University of California, Los Angeles}
\email{zesenhuang@g.ucla.edu}

\author[0000-0002-2381-3106]{Marco Velli}
\affiliation{Department of Earth, Planetary, and Space Sciences, University of California, Los Angeles}
\email{mvelli@ucla.edu}

% \author[0000-0002-2582-7085]{Chen Shi}
% \affiliation{Department of Physics, Auburn University}
% \email{chs0090@auburn.edu}

\author[0009-0009-0162-2067]{Yuliang Ding}
\affiliation{Department of Earth, Planetary, and Space Sciences, University of California, Los Angeles}
\email{dingyl@g.ucla.edu}

\begin{abstract}

We present a solitary Alfv\'en wave model that exhibits nontrivial three-dimensional twisting of open magnetic field lines while preserving constant $|B|$. Embedded rotational discontinuities sharply deflect the otherwise uniform field lines, producing localized, large-amplitude field reversals in one-dimensional profiles that closely resemble the ``switchbacks'' observed by the Parker Solar Probe in the inner heliosphere. This indicates that switchbacks, as seen in one-dimensional spacecraft time series, arise from traversals through strongly curved segments of open magnetic field lines.

\end{abstract}


\section{Introduction}\label{sec:intro}

One of the most unexpected discoveries of the Parker Solar Probe (PSP) \citep{fox_solar_2016} is the ubiquity of magnetic ``switchbacks'' in the pristine solar wind \citep{raouafi_parker_2023}. Switchbacks are magnetic field reversals with nearly constant magnetic field magnitude, often accompanied by one-sided anti-sunward proton jets that exhibit near-perfect Alfv\'enic correlation \citep{bale_highly_2019,gosling_one-sided_2009}. Close to the Sun, switchbacks are typically solitary, whereas further away, they tend to grow in size and merge. Importantly, electron pitch angle distributions (PADs) show field-aligned anti-sunward electron strahls both inside and outside the field reversal regions \citep{kasper_alfvenic_2019,huang_solitary_2025}, suggesting open field-line topology. Thus, a substantial portion of switchbacks resemble spherically polarized Alfv\'en waves (SPAWs), which are exact nonlinear solutions to the ideal MHD equations \citep{goldstein_theory_1974,barnes_nonexistence_1976}.

The key to understanding the physics of switchbacks is to construct a solitary model, which is, however, mathematically challenging. The difficulty stems from constructing a divergence-free, constant-magnitude vector field with open field-line topology. Several studies have attempted to construct switchback models, but they either contained topologically closed regions \citep{tenerani_alfvenic_2020,shi_analytic_2024} or failed to reproduce the solitary behavior \citep{squire_construction_2022,shoda_turbulent_2021}. In a companion study \cite{huang_solitary_2025}, we constructed a numerical model of a solitary Alfv\'en wave packet (switchback) with open field-line topology, termed \emph{Alfv\'enon}. Direct MHD simulations demonstrate remarkable stability, validating it as an exact solution to the ideal MHD equations.  In this study, we attempt to address the question ``What are switchbacks?'' by examining the field-line structure of the switchback model.

This paper is organized as follows. In Section~\ref{sec:model}, we layout the governing equations for solitary Alfv\'en waves and describe the general properties of the switchback model. Section~\ref{sec:results} examines the field-line structure of the model and its correspondence to one-dimensional observations of switchbacks. Section~\ref{sec:discussion} presents our discussion. Section~\ref{sec:conclusions} concludes the study.

\section{The Switchback Model}\label{sec:model}

In the companion study \citep{huang_solitary_2025}, we solved the ideal MHD equations:
\begin{align}
    \frac{\partial \rho}{\partial t} + \nabla\cdot (\rho \vec u) &= 0,\label{eq:mhd1}\\
    \rho\left[\frac{\partial \vec u}{\partial t} + (\vec u \cdot \nabla) \vec u \right] &= -\nabla \left(p+\frac{B^2}{2\mu_0}\right) + \frac{1}{\mu_0} (\vec B\cdot \nabla) \vec B,\label{eq:mhd2}\\
    \frac{\partial \vec B}{\partial t} &= \nabla \times (\vec u \times \vec B),\label{eq:mhd3}\\
    p \rho^{-\gamma} &= \mathrm{const},\label{eq:mhd4}\\
    \nabla \cdot \vec B &= 0.\label{eq:mhd5}
\end{align}
Under the assumptions ($\mathcal{A}1$) constant $|\vec B|$, ($\mathcal{A}2$) constant $\rho$, and ($\mathcal{A}3$) constant $p$, we seek \emph{solitary Alfv\'enic} solutions of the form $\vec u (\vec r) = \vec u_0 + \vec u_1(\vec r)$ and $\vec b(\vec r) = \vec b_0 + \vec b_1(\vec r)$, where $\vec b = \vec B/\sqrt{\rho \mu_0}$. The solitary condition requires $\vec b_0 = \vec b(\infty)$ and hence $|\vec b| = |\vec b_0| = |\vec b_0 + \vec b_1|$. The Alfv\'enic condition dictates $\vec u_1 = \pm \vec b_1$. Since the magnetic field is Galilean-invariant, we work in the frame where $\vec u_0 = 0$, yielding:
\begin{align}
    \frac{\partial \vec u_1}{\partial t}  &=  \vec b_0 \cdot \nabla \vec b_1, \label{eq:alfven1}\\
    \frac{\partial \vec b_1}{\partial t} &= \vec b_0 \cdot \nabla\vec u_1. \label{eq:alfven2}
\end{align}
Combining \eqref{eq:alfven1} and \eqref{eq:alfven2} gives the wave equations:
\begin{align}
    \frac{\partial^2 \vec u_1}{\partial t^2}  &=  (\vec b_0 \cdot \nabla)^2 \vec u_1, \label{eq:wave1}\\
    \frac{\partial^2 \vec b_1}{\partial t^2}  &=  (\vec b_0 \cdot \nabla)^2 \vec b_1, \label{eq:wave2}
\end{align}
which describe wave solutions $\vec b_1(\vec r \pm \vec b_0 t)$, where $\vec b_0$ is the \emph{unperturbed} field and $\vec b_1$ is the \emph{perturbation}. Analogous to field-line perturbations in shear Alfv\'en waves \citep{alfven_existence_1942}, $\vec b_1$ represents a three-dimensional non-trivial twisting of magnetic field lines that preserves field-line density (constant $|B|$) and maintains open field-line topology. In the mean time, the correlation $\vec u_1 = \pm \vec b_1$ determines the direction of propagation w.r.t. $\vec b_0$. 

\begin{figure}[!bthp]
    \centering
    \includegraphics[width=\columnwidth]{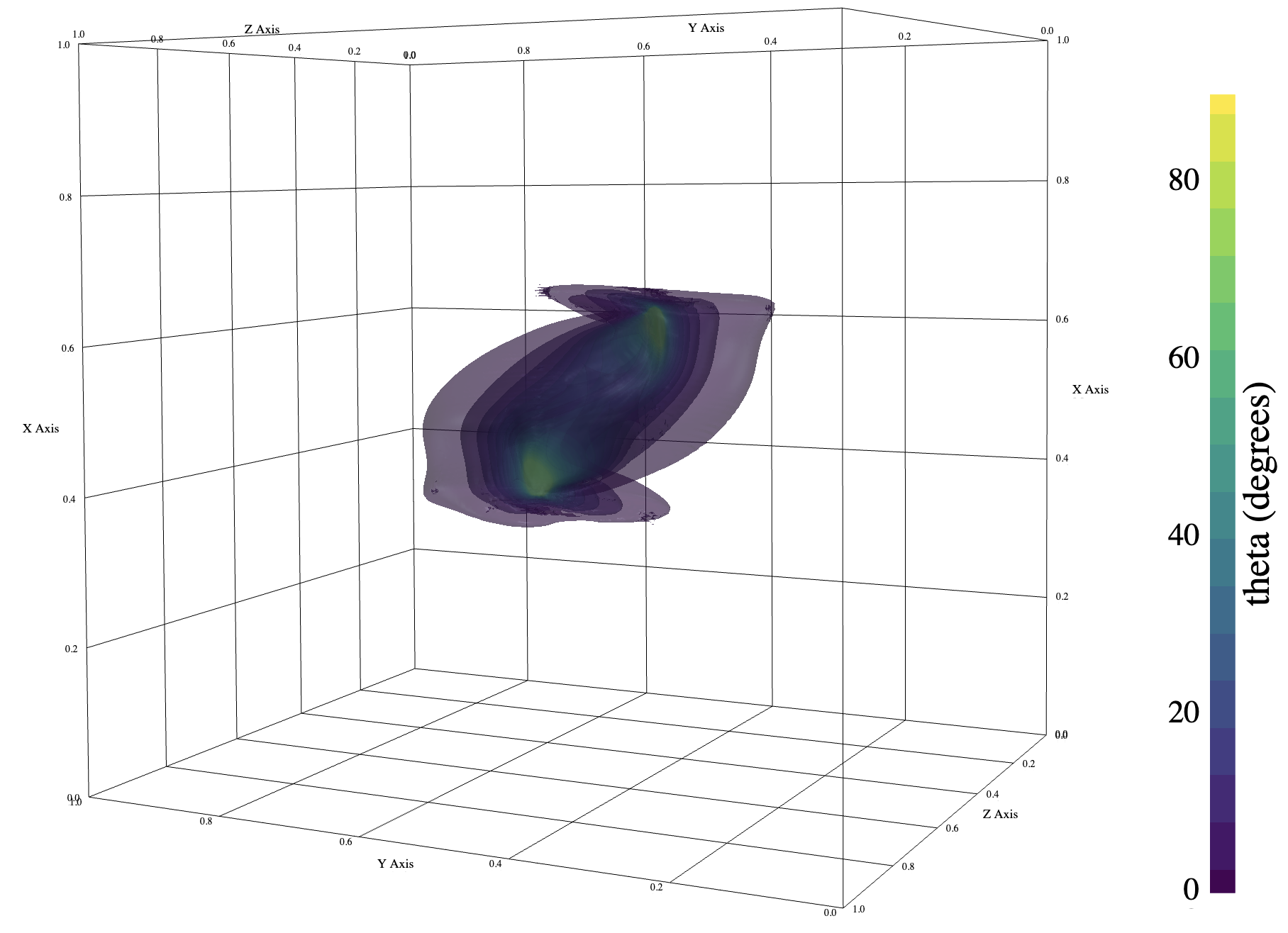}
    \caption{Contour of $\theta(x,y,z)$ of the switchback model.}
    \label{fig:model}
\end{figure}

To construct a numerical model of $\vec b(x,y,z)$, we introduced a convergent iterative algorithm \citep{huang_solitary_2025}:
\begin{eqnarray}
    \mathcal{H}:\;\vec F \;\mapsto\; \vec F' = \frac{\vec F - \nabla\varphi}{\,|\vec F - \nabla\varphi|}
\end{eqnarray}
where $\nabla\varphi$ is obtained with Helmholtz-Hodge decomposition of $\vec F$. After enough iterations, it converges to a stable divergence free vector field $\vec G_A$:
\begin{eqnarray}
    \vec F_n = \mathcal{H}^n(\vec F_0),\quad
    \vec G_n = \vec F_n - \nabla\varphi_n
    \;\overset{n\to\infty}{\longrightarrow}\; \vec G_A,    
\end{eqnarray}
where $\vec F_0(x,y,z)$ is defined as:
\begin{eqnarray}
\vec F_0 &= \vec B_0 
+ A\!\left[\cos(\phi)\,\hat{y} + \sin(\phi)\,\hat{z}\right]
        \exp\!\left(-\frac{\Delta r^{2}}{2\sigma^{2}}\right)
\end{eqnarray}
where $\vec B_0 = (1,0,0),\;\phi = 2\pi k_x x,\; k_x = 4,\; \Delta r = \sqrt{(x-0.5)^2 + (y-0.5)^2 + (z-0.5)^2}$ and $\sigma = \tfrac{1}{30}$. Here we chose $A=100.0$, and the algorithm yielded a non-trivial solitary vector field $\vec{G}_A$ exhibiting large-amplitude local field reversals with open field-line topology in a $256^3$ grid ($|\vec G_A|\simeq 1$, $\nabla \cdot \vec G_A = 0$). Its structure closely resembles the switchbacks observed \textit{in situ} in the upper corona by PSP \citep{kasper_alfvenic_2019,bale_highly_2019,kasper_parker_2021,huang_solitary_2025}. We will henceforth refer to $\vec{G}_A$ as the switchback model.

Figure~\ref{fig:model} shows the three-dimensional contour of $\theta(x,y,z) = \cos^{-1}[\vec{B}(x,y,z) \cdot \vec{B}_0]$, where $\vec{B}(x,y,z)$ is the switchback model $\vec{G}_A(x,y,z)$, and $\vec{B}_0 = (1,0,0)$ represents the asymptotic unperturbed field value near the edges of the $256^3$ computational domain. The coordinates $x$, $y$, and $z$ each span the range $[0,1]$. The perturbed region of $\vec{G}$ is concentrated near the center of the domain, where two strongly perturbed regions exhibit $\theta \gtrsim 90\degree$, indicating local field reversals. Outside this perturbed region, $\vec{G}_A$ asymptotically approaches $\vec{B}_0$. An interactive version of Fig.~\ref{fig:model} is available in the supplementary materials.

\section{Switchback Structure}\label{sec:results}

\begin{figure*}[!bthp]
    \centering
    \includegraphics[width = \textwidth]{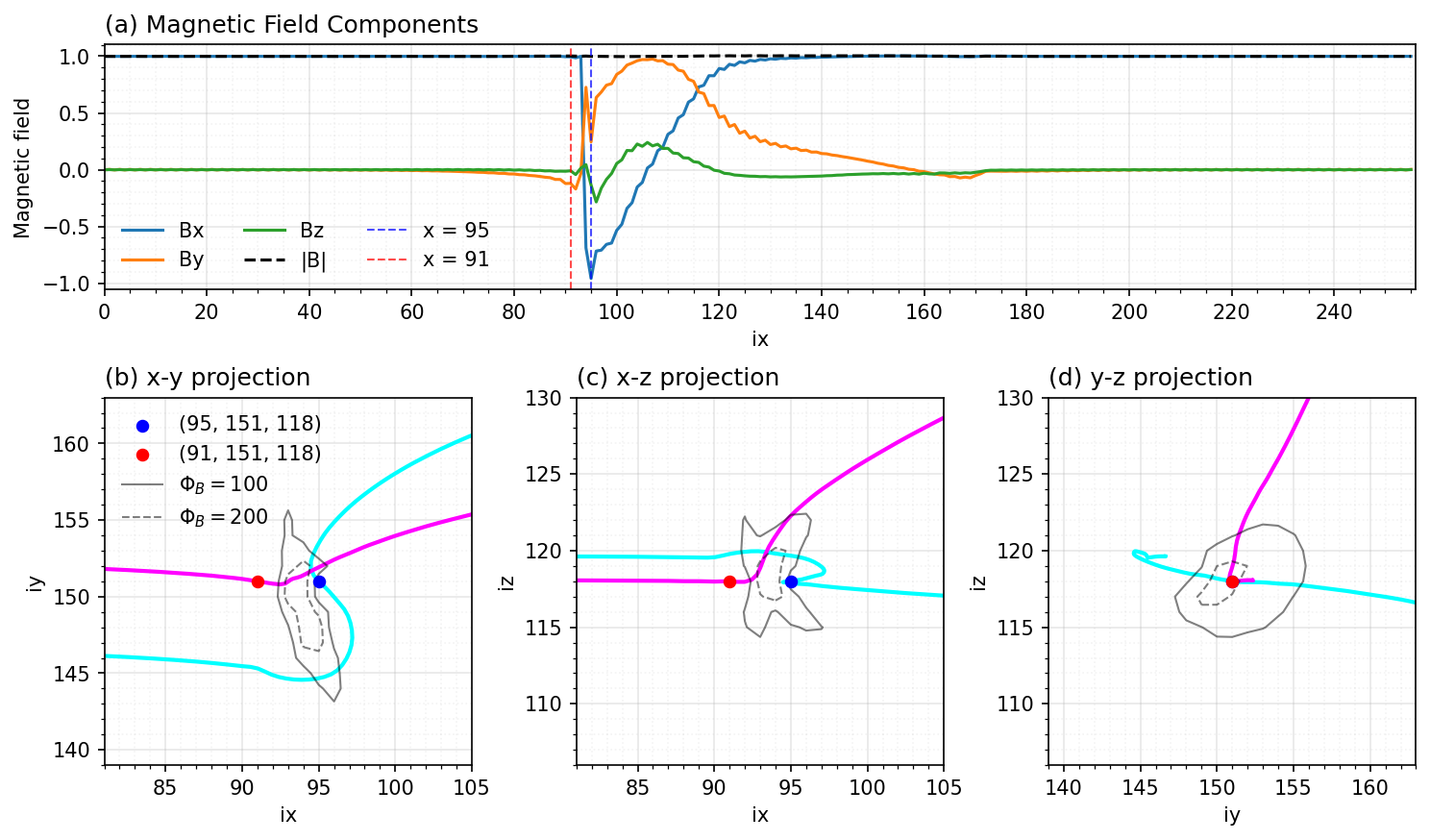}
    \caption{
        Field-line structure of a switchback. (a) 1D profile at iy=151 and iz=118. (b-d) x-y/x-z/y-z projection of the field-lines passing (91,151,118) and (95,151,118). Contours show equivalue lines of $Phi_B=100,200$.
    }
    \label{fig:1D_profile}
\end{figure*}

\begin{figure}[!bthp]
    \centering
    \includegraphics[width=\columnwidth]{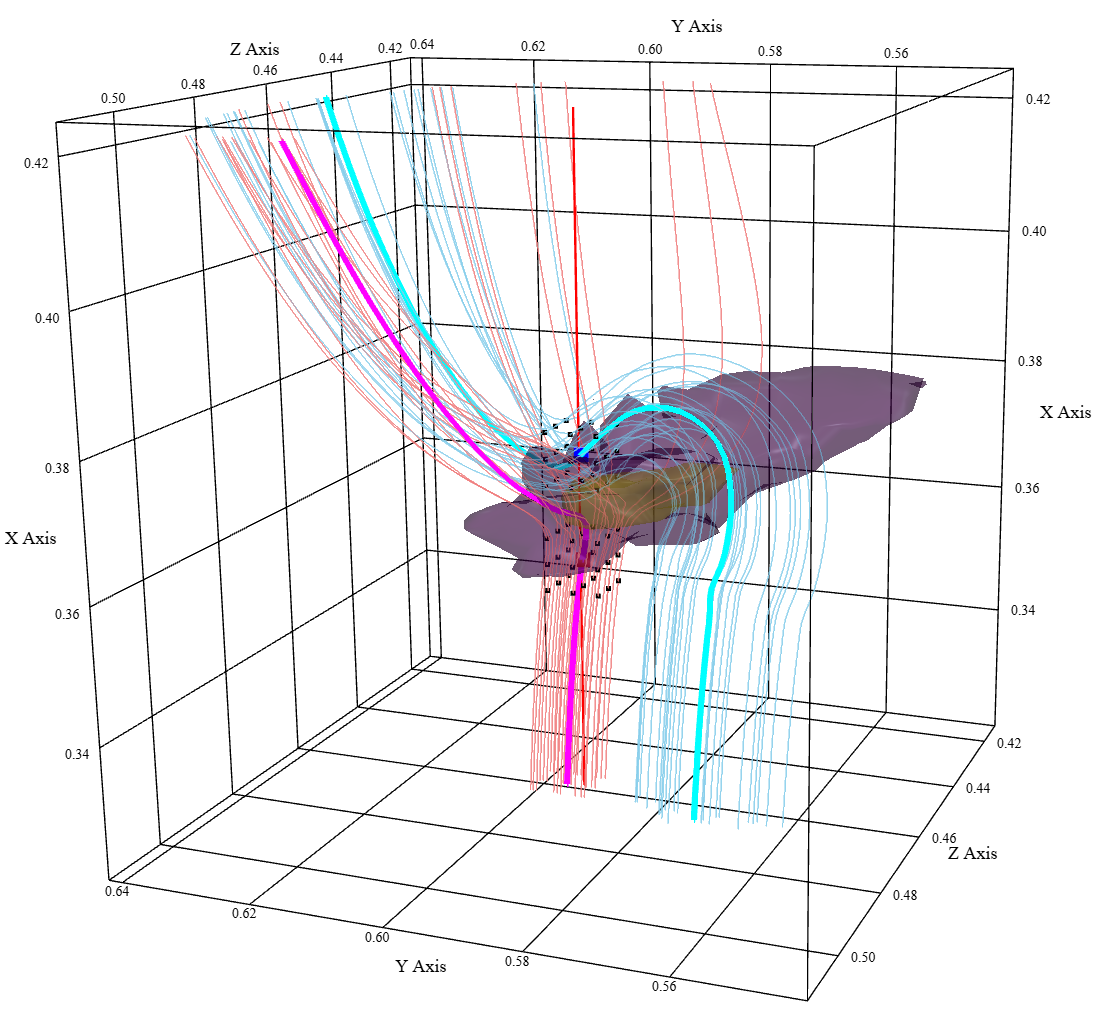}
        \caption{Contour surfaces of $\Phi_B(x,y,z)$ at values 100 (purple) and 200 (yellow), with magnetic field lines overlaid. The red line indicates the x-direction profile through $(95,151,118)$. Magenta and cyan curves show representative field lines passing through $(91,151,118)$ (red) and $(95,151,118)$ (blue), respectively. Small black dots mark the 26 neighboring grid points around each location, with their field lines shown in light red and light blue respectively.}
    \label{fig:switchback}
\end{figure}

To understand the structure of the switchback model, we examine the three-dimensional field $\vec B(x,y,z)$. The deflection angle $\theta(x,y,z)$ reaches a maximum of $163.713\degree$ at grid indices $(95, 151, 118)$ and $(161, 105, 118)$, corresponding to the centers of the two strongly perturbed regions visible in Fig.~\ref{fig:model}. Figure~\ref{fig:1D_profile}(a) shows the one-dimensional profile of $\vec B$ along the $x$-axis passing through $(95, 151, 118)$. A sharp reversal of $B_x$ occurs at $ix = 95$, while $|B|$ remains perfectly constant, suggesting a rotational discontinuity. This characteristic is consistent with \textit{in situ} observations, where switchback boundaries are frequently identified as sharp rotational discontinuities \citep{huang_solitary_2025}.

Alfv\'en waves are driven by the magnetic tension force $(\vec B \cdot \nabla)\vec B$. The right-hand side of Eq.~\eqref{eq:alfven1} describes the \emph{restoring force} on $\vec u_1$ arising from local field-line curvature. Thus, 
\begin{equation}
    \Phi_B(x,y,z) = |(\vec B_0\cdot \nabla)\vec B|
    \label{eq:phi_B}
\end{equation}
serves as a convenient indicator of local field-line curvature. In \cite{huang_solitary_2025}, we showed the contour of $\Phi_B$, whose values range from 0 to over 200. Only two zones exhibit $\Phi_B > 100$, located in the two most strongly perturbed regions in Fig.~\ref{fig:model}. Here we compute the isosurfaces of $\Phi_B = 100$ and $\Phi_B = 200$ near $(95, 151, 118)$, shown in Fig.~\ref{fig:switchback}.

The red line marks the one-dimensional profile shown in Fig.~\ref{fig:1D_profile}(a). To visualize the switchback structure, we trace two representative field lines passing through $P_1(91, 151, 118)$ and $P_2(95, 151, 118)$, shown in magenta and cyan, respectively. Both field lines enter the computational domain at the $x=0$ plane, initially aligned with the $+x$ direction. Figure~\ref{fig:1D_profile}(b-d) presents the two-dimensional projections of these field lines onto the $x$-$y$, $x$-$z$, and $y$-$z$ planes. As the field lines approach the strongly perturbed zones, they undergo significant deflection in regions where $\Phi_B$ is large, which effectively act as deflectors. The two groups of 26 neighboring grid points surrounding $P_1$ and $P_2$ are marked as small black dots in Fig.~\ref{fig:switchback}. Field lines passing through these neighboring points (shown in light red and light blue) similarly exhibit deflection as they encounter the high-$\Phi_B$ region, demonstrating the localized nature of the field-line distortion.

The ``switchback'' in Fig.~\ref{fig:1D_profile} therefore arises from traversing the strongly perturbed region where field lines exhibit significant curvature. The drastically varying field-line curvature manifests as a sharp field direction reversal in the one-dimensional profile. Notably, the $\Phi_B$ contours effectively visualize the three-dimensional structure of this rotational discontinuity. Any one-dimensional cut through this region will exhibit sharp field direction reversals, i.e. ``switchbacks''.

\section{Discussions}\label{sec:discussion}

\begin{figure}
    \centering
    \includegraphics[width=\columnwidth]{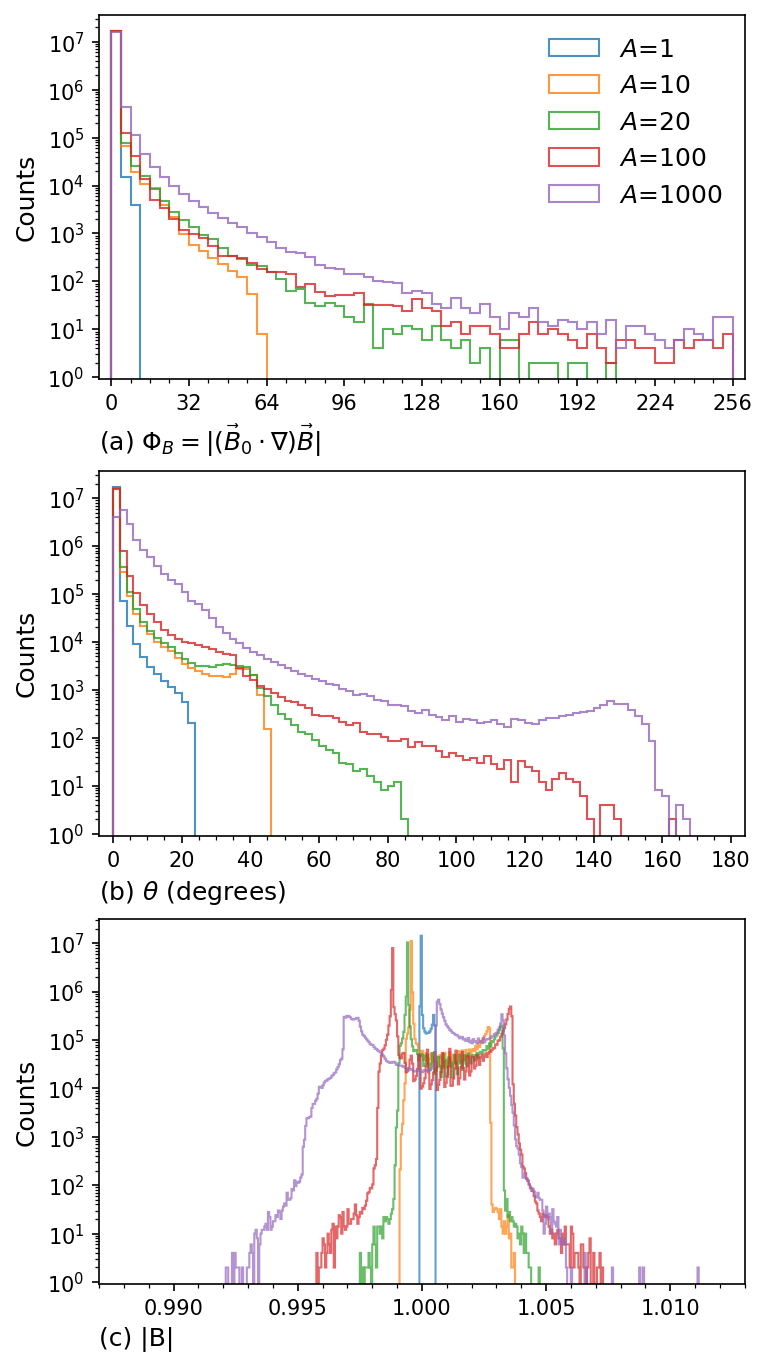}
    \caption{Distribution of (a) $\Phi_B$, (b) $\theta$, and (c) $|B|$ for different initial amplitudes $A$ in $\vec F_0$.}
    \label{fig:hist}
\end{figure}

% \subsection{Rotational Discontinuities}
When computing $\Phi_B = |(\vec B_0\cdot \nabla)\vec B|$, we employ the symmetric difference quotient for the spatial derivative. Therefore, the theoretical maximum value of $\Phi_B$ is $|2\vec B_0|/(2\Delta x) = 256$, where $\Delta x = 1/256$ is the grid spacing. In the presence of a strong rotational discontinuity, the local $\Phi_B$ exhibits extreme values, making it a reliable indicator of rotational discontinuities in the solution. 

The structure of the switchback model depends strongly on the initial condition $\vec F_0$, particularly the amplitude parameter $A$. To explore the behavior of the convergent algorithm, we constructed solutions for various amplitudes $A$ in $\vec F_0$ while keeping all other parameters fixed. For each value of $A$, we performed 200 iterations to obtain the final converged field $\vec G_{A}$. We selected five representative amplitudes: $A = 1, 10, 20, 100, 1000$. The resulting histograms of $\Phi_B$, $\theta$ and $|B|$ are shown in Figure~\ref{fig:hist}.

One particular interesting finding is that the solutions have dramatically different behaviors for $A \lesssim 20$ and $A\gtrsim 20$. For $A<20$, $\Phi_B$ values are small, and hence the field line twistings in the solutions are smooth, resulting in small $\theta$ values. Starting from $A\simeq 20$, $\Phi_B$ start to exhibit extreme values, indicating the existence of rotational discontinuities. Consequently, the distribution of $\theta$ extend to large angles, thereby forming switchbacks. For all amplitudes, $|B|$ remains remarkably constant, albeit showing increasing deviation. 

Notably, for $A\gtrsim 20$, deviations in $|B|$ form highly localized defects that typically occupy only a few grid points (often just two points due to symmetry, as seen in the $A=1000$ curve in panel c). As illustrated in Fig.~\ref{fig:1D_profile}(a), the switchback appears as a sharp discontinuity surrounded by high-$k$ perturbations. This structure arises from the interplay between the \emph{Gibbs phenomenon} and the convergent algorithm: the Gibbs phenomenon inevitably produces approximately 9\% overshoot/undershoot at discontinuities while converging almost everywhere else, whereas the algorithm attempts to enforce constant $|B|$. The combined effect results in small local defects in $|B|$ accompanied by high-$k$ perturbations near the discontinuity.

Our results indicate that rotational discontinuities are not required for the formation of small-amplitude solitary Alfv\'en waves. Instead, the system likely admits smooth field-line twisting only up to a critical angle $\theta_c$. Beyond this threshold, continuous solutions cease to exist, and rotational discontinuities become necessary to sustain perturbations with $\theta > \theta_c$. This behavior is analogous to the loss of smooth torsional equilibria in Kirchhoff rod theory and resembles the onset of the fire-hose instability, where the system undergoes an abrupt transition to a qualitatively different configuration. A similar twist-to-writhe transition is well documented in DNA mechanics, where excessive torsional stress leads to supercoiling once smooth twisting can no longer be sustained.

Finally, we note that this behavior is contingent upon strict enforcement of the constant $|B|$ constraint. If higher compressibility is permitted, smooth solutions with large $\theta$ may be achievable. This is partly supported by spacecraft observations: switchbacks are typically sharp in the low-$\beta$ environment of the upper solar corona, where $|B|$ remains remarkably constant. Farther out in the heliosphere, as plasma $\beta$ increases and the strict constancy of $|B|$ is relaxed, switchbacks become larger both in spatial extent and amplitude. Future work will investigate this phenomenon in greater detail.

\section{Conclusions}\label{sec:conclusions}

The answer to the question raised in the title is straightforward: switchbacks, as observed in spacecraft time series data, are one-dimensional traversals through nontrivial twisting of open magnetic field lines. Under the strict constant $|B|$ condition, our model suggests that switchbacks are associated with three-dimensional rotational discontinuities that strongly deflect magnetic field lines, consistent with PSP \textit{in situ} observations in the magnetically dominant upper solar corona. This behavior is indicative of intriguing ``elastic'' properties of magnetic field lines, analogous to the twist-to-writhe transition in DNA mechanics.

\begin{acknowledgments}
Z.H. thanks Melvyn Goldstein for stimulating discussions. This work is supported by NASA HTMS 80NSSC20K1275 and NASA AIAH 80NSSC26K0128.
\end{acknowledgments}

% \clearpage

% \appendix

\bibliography{references}{}
\bibliographystyle{aasjournalv7}

\end{document}